\documentclass[twocolumn,showpacs,preprintnumbers,amsmath,amssymb]{revtex4}
\usepackage{graphicx}% Include figure files
\usepackage{dcolumn}% Align table columns on decimal point
\usepackage{bm}
\begin{document}
%\preprint{APS/123-QED}
%\usepackage{epsfig}
\title{Quantum computation with un-tunable couplings}
\author{Xingxiang Zhou,$^{\dagger,1}$Zheng-Wei Zhou,$^{\ddagger,2}$
     Guang-Can Guo,$^2$ and Marc J. Feldman$^1$}
%\author{Xingxiang Zhou}
%\email{xizhou@ece.rochester.edu}
%\author{$^{,1}$Zheng-Wei Zhou}
%\email{zwzhou@ustc.edu.cn}
%\author{$^{,2}$ Guang-Can Guo$^2$}
%\author{$^2$ Marc J. Feldman$^1$}
\affiliation{ $^1$Superconducting Digital Electronics Lab,
Electrical and Computer Engineering Department, University of
Rochester, Rochester, NY 14627, USA  \\ $^2$Key Laboratory of
Quantum Information, University of Science and Technology of
China, Chinese Academy of Sciences, Hefei, Anhui, China, 230026.}
\date{\today}

\begin{abstract}
Most quantum computer realizations require the ability to apply
local fields and tune the couplings between qubits, in order to
realize single bit and two bit gates which are necessary for
universal quantum computation. We present a scheme to remove the
necessity of switching the couplings between qubits for two bit
gates, which are more costly in many cases. Our strategy is to
compute in and out of carefully designed interaction free
subspaces analogous to decoherence free subspaces, which allows us
to effectively turn off and turn on the interactions between the
encoded qubits. We give two examples to show how universal
quantum computation is realized in our scheme with local
manipulations to physical qubits only, for both diagonal and off
diagonal interactions.
\end{abstract}


\maketitle

%Despite its promise to solve certain problems fundamentally faster
%than any classical computing machines, the construction of a large
%scale quantum computer is far beyond our current available
%technology. Much of the difficulty lies in the necessity of both
%single bit and two bit operations, following the theoretical
%construction of arbitrary unitary gates from these elements
%\cite{ref:Universal}.
Quantum computation is generally formulated in terms of a
collection of qubits subject to a sequence of single and two bit
operations \cite{ref:Universal}. This implies that the effective
local fields applied to individual qubits, and the couplings
between the qubits, are variable functions subject to external
control. In many cases, two bit operations, whose implementation
depends on certain interactions between qubits, are more difficult
than single bit gates. They can require more sophisticated
manipulations, therefore may take a longer time and cause stronger
decoherence. %Following the the language of quantum circuits, which
%consist of a sequence of single and two bit operations, most
%quantum computer proposals require the ability to vary (in the
%simplest case just turn on and off) the couplings between qubits.
This usually results from the requirement to vary (in the simplest
case just switch on and off) the couplings between qubits, which
is not always possible, or easy to realize. One such example is
quantum computing with Josephson junction devices, both charge and
flux type
\cite{ref:Charge1,ref:Charge2,ref:Flux1,ref:Flux2,ref:Flux3}. In
this case, the coupling between qubits is most naturally realized
with a hard wired capacitor or inductor, whose value is fixed by
the fabrication and cannot be tuned during the computation. The
superconducting quantum computing community has been working hard
to devise variable coupling schemes
\cite{ref:Flux2,ref:Flux-Switch,ref:Charge-Switch1,
ref:Charge-Switch2}, but it is generally agreed that none of these
proposed switches is completely satisfactory
\cite{ref:Charge-Switch2}. Most of them
\cite{ref:Flux2,ref:Flux-Switch} require external controls, thus
are likely to be major decoherence sources. Others were designed
to avoid such external controls, but may suffer other problems,
for instance the number of qubits that can be incorporated into
the system can be limited \cite{ref:Charge-Switch1,
ref:Charge-Switch2}, which is at odd with the supposed scalability
of a solid state quantum computer.
%In ref.
%\cite{ref:Charge-Switch1, ref:Charge-Switch2}, a method to achieve
%controllable couplings by using a LC resonator ``bus'' was
%presented. This approach has the advantage that no external
%control is needed, but the number of qubits that can be attached
%to the ``bus'' is limited, which is at odd with the supposed
%scalability of a solid state quantum
%computer. %Other examples of hard-to-tune couplings  include
%...\cite{ref:Others}.

An always on and un-tunable coupling causes certain problems for
quantum computation, depending on the particular form of the
interaction. If the interaction Hamiltonian is diagonal in the
computational basis, each qubit state will gain additional phases
depending on the states of the qubits to which it is coupled, even
in the idle mode. It is then necessary to keep track of these
phases, or suppress them by repeated refocusing pulses like those
used in NMR, which requires high precisions and complicates the
operation \cite{ref:Charge-Switch1, ref:Charge-Switch2}. The
situation is more serious in the case of off diagonal
interactions, because these interactions will cause the states of
the qubits to propagate, which results in errors. It is then
necessary to devise methods to avoid these problems. Even if the
couplings can be tuned, a scheme which allows to compute without
switching the couplings is very useful, because it simplifies the
operation drastically and is likely to help reduce decoherence.
This simplified computing scheme, in which the necessity of
switching the couplings between qubits is removed, is the goal of
this work, and we attack this problem by computing in carefully
designed subspaces analogous to decoherence free subspaces.

Let us first explain our approach intuitively. One of the
strategies that people came up with in the effort to fight against
decoherence is to compute in the so called ``decoherence free
subspace'' (DFS), in which the state of the system (a logical bit
consisting of several physical qubits) is unaffected by the
environment even though they are always coupled \cite{ref:DFS}.
DFS exists in the case of ``collective docoherence'', {\it i.e.},
when the involved qubits couple to the same mode of the
environment. Now imagine that we replace the environment with
another collection of qubits. Obviously, we expect analogous
subspaces of the two coupled collections of qubits exist, as long
as certain conditions similar to those for DFS are satisfied. To
be more specific, states that are annihilated by the interaction
Hamiltonian will not evolve because of the coupling. If we stay in
these subspaces, we can then prevent the (encoded) qubits from
affecting each other and operate on the individual logical bits as
if they were not coupled to other bits. On the other hand, when we
do want the (encoded) bits to interact for two bit operations, we
simply drive them out of these subspaces. Therefore, we can
effectively turn off and turn on the interaction between the
encoded bits by staying in and getting out of these subspaces. We
see that even though the subspaces we discussed rely on the same
algebraic properties with DFS, we are using them for a different
purpose, and we do not intend to stay in them throughout the
computation, as opposed to DFS. Another important difference is
%that in the case of DFS, it is assumed that the environment and
%its coupling strength with the qubits are un-controllable.
that DFS is a ``passive'' method of protecting the quantum
information. It is assumed that the environment and its coupling
strength with the qubits are un-controllable. On the contrary, in
our case we can certainly manipulate both parties involved and put
them in or out of their related subspaces. We can choose the way
and the strength with which the qubits are coupled, either in
fabrication or during initialization of the system, though the
couplings cannot be tuned throughout the computation. Also, the
number of qubits that can be put in the DFS is limited by the
wavelength of the environmental modes, whereas our scheme is very
extensible. Because of these important distinctions, we call this
new concept the ``Interaction free subspace'' (IFS).

In the following we discuss in detail how our scheme can be used
for both diagonal and off diagonal interactions. A general model
for a quantum computer is described by the following Hamiltonian:
\begin{equation}
H=-\sum_i \vec{f}_i\cdot
\vec{\sigma}_i+\sum_{i<j}\sum_{\alpha,\beta}J_{ij}^{\alpha
\beta}\sigma_i^\alpha\sigma_j^\beta,
\end{equation}
where $\vec{f}_i$ is the effective local field applied to
individual qubits, $\sigma^\alpha$, $\sigma^\beta$ are the pauli
matrices, and $J$ is the coupling strength. As we discussed
before, we are interested in situations in which local operations
are easy and fast to implement, while two bit operations are hard
and slow. We assume that local resources are ``free'', {\it i.e.},
strong local pulse fields $\vec{f}_i$ can be applied and single
bit gates are instantaneous. We then only count the time when the
interaction is on ({\it i.e.}, when the encoded bits sit out of
the IFS). we assume that the values of $J_{ij}$, {\it i.e.}, the
coupling strength between qubits can be chosen in the fabrication
or initialization of the system, but cannot be tuned during the
computation. (The dependence on $\alpha, \beta$ is determined by
the nature of the physical interaction.) When the interaction is
on, we occasionally need to apply local gates to the individual
qubits involved, and we assume that the field used for these local
gates is so strong that the local operation is not distorted by
the interactions.

\begin{figure}[h]
%    \centering
%    \epsfig{file=aa1.eps, width=2.3in, height=1.2in}
    \includegraphics[width=3in, height=1.2in]{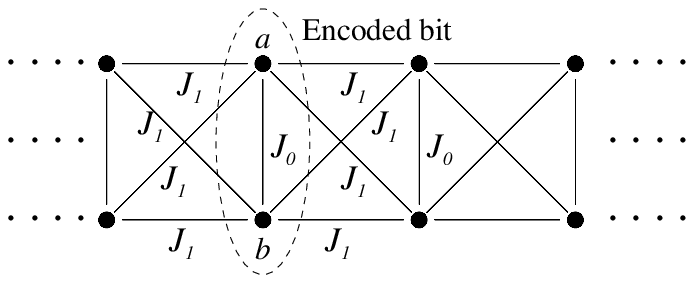}
    \caption{Architecture of the quantum computer for diagonal interactions.
    Each dot is a physical qubit and the lines represent couplings between
    qubits. Two qubits (a, b) connected by a vertical line is an encoded qubit.}
    \label{fig:Diag}
\end{figure}

We first consider the case of diagonal interactions, {\it i.e.},
when the interaction between two (physical) qubits takes the form
$J_{12}^z\sigma_1^z\cdot\sigma_2^z$ (the Ising interaction). In
this case, two physical qubits per logical bit can fulfill our
needs. As shown in Fig. \ref{fig:Diag} a possible architecture of
the quantum computer in this case is a one dimensional array
consisting of encoded qubits, which are two physical qubits ($a$,
$b$ in the figure) coupled with strength $J_0$. All physical
qubits in neighboring logical bits are coupled with the same
strength $J_1$, which is not necessarily different from $J_0$. Our
codes for the IFS are simply
$|0\rangle=|\uparrow_a\downarrow_b\rangle$ and
$|1\rangle=|\downarrow_a\uparrow_b\rangle$. Indeed, for two
neighboring encoded bits, the interaction Hamiltonian is
$H_{int}=J_1(\sigma_{1a}^z+\sigma_{1b}^z)\cdot
(\sigma_{2a}^z+\sigma_{2b}^z)$, which annihilates these two
states. In addition, these two states are degenerate under the
self Hamiltonian $J_0\sigma_a^z\cdot \sigma_b^z$. Therefore, if we
store information in these states, no evolution whatsoever is
present. There is thus no need to keep track of any phases.
Actually, in order to avoid interactions between the encoded
qubits, it suffices to keep half of the qubits (all odd or even
numbered ones) in the IFS. However, as will be seen it is
inevitable to get out of the IFS even when single bit operations
are performed on the encoded qubits, since the only resources we
have are single (physical) bit rotations. We thus keep all the
logical bits in the IFS during the idle mode.

Now we discuss how universal computation can be realized on these
encoded qubits, using local operations to the physical qubits
only. Suppose we are operating on a particular encoded bit
(meaning applying single bit gates on its qubit $a$ and $b$).
Since its neighboring bits are in IFS, we can work on the
particular bit as if it were decoupled from the rest of the
system. First, we notice that single bit gates can be decomposed
into arbitrary rotations around the $z$ axis and the Hadamard gate
\begin{equation}
\textsf{H} =\frac{1}{\sqrt{2}}\left(
\begin{array}{cc}
1 & 1 \\
1 & -1 \\
\end{array}
\right ). \label{eq:Hadamard}
\end{equation}
To induce a rotation around the $z$ axis, all we have to do is to
break the degeneracy between $|0\rangle$ and $|1\rangle$, which
can be done by applying a local field in the $z$ direction to bit
$a$ (or $b$) for a certain amount of time. A Hadamard gate on the
bases $|0\rangle$, $|1\rangle$ is more complex, but as can be
easily verified it can be realized with the gate sequence $CNOT(a,
b)\cdot \textsf{H}_a\cdot CNOT(a, b)$, where $CNOT(a, b)$ is a
$CNOT$ gate between $a$, $b$ with $a$ as the control bit, and
$\textsf{H}_a$ is a Hadamard on bit $a$. $\textsf{H}_a$ is readily
realizable, while $CNOT(a, b)$ can be done by sandwiching a
$CPHASE$ gate between $a$ and $b$ with $\textsf{H}_b$. The
$CPHASE$ gate is easy to implement with the Ising interaction and
local phase gates: $CPHASE=e^{-i\sigma_a^z
\sigma_b^z\pi/4}e^{i\sigma_a^z\pi/4}e^{i\sigma_b^z\pi/4}$.
Altogether, a Hadamard on the code states requires $9$ local gates
and $2$ interaction periods, which takes a time $\pi/2J_0$ (local
gates are assumed instantaneous). We note that it is necessary to
get out of the code space $|\uparrow_a\downarrow_b\rangle$,
$|\downarrow_a\uparrow_b\rangle$ in order to realize the Hadamard.
This is unavoidable in the current model, since the only allowed
resources are local unitaries. As a result, single bit gates on
the encoded qubits cannot be performed simultaneously to
neighboring bits. But a ``half parallel'' operation mode is still
allowed, in which all odd or even numbered logical bits are
operated on at the same time. This restriction can be removed by
exploring more complex encoding schemes (see the example for off
diagonal interactions below), but more resources (more physical
qubits per logical bit) are required.

We still need to show how two bit gates can be realized. For this
purpose, we need drive the involved (neighboring) bits out of the
IFS, let them interact for some time, then drive them back. The
first step is to apply a local gate to flip the state of $b$,
which changes the two code states to
$|\uparrow_a\uparrow_b\rangle$ and $|\downarrow_a
\downarrow_b\rangle$. Note these two states are eigenstates of
$\Sigma^z/2=(\sigma_a^z+\sigma_b^z)/2$ with eigenvalues $\pm 1$. %The
%dynamics of $\Sigma^z$ on these states is thus the same with that
%of $\sigma_z$ on $|\uparrow\rangle$ and $|\downarrow\rangle$.
%in an Ising model, {\it i.e.}, the Hamiltonian of two neighboring bits
%subject to local fields in the $z$ direction is
%$H_{12}=B_1^z\Sigma_1^z+ B_2^z\Sigma_2^z
%+J_1\Sigma_1^z\Sigma_2^z/4$.
It is then straight forward that a $CPHASE$ gate between 1, 2 (on
$|\uparrow_a\uparrow_b\rangle$ and $|\downarrow_a
\downarrow_b\rangle$) can be realized with the sequence
$e^{-i(\Sigma_1^z/2) (\Sigma_2^z/2)\pi/4} e^{i(\Sigma_1^z/2)\pi/4}
e^{i(\Sigma_2^z/2)\pi/4}$, or $e^{-i(\sigma_{1a}^z
+\sigma_{1b}^z)(\sigma_{2a}^z +\sigma_{2b}^z)\pi/16}
e^{i\sigma_{1a}^z\pi/8} e^{i\sigma_{1b}^z\pi/8}
e^{i\sigma_{2a}^z\pi/8} e^{i\sigma_{2b}^z\pi/8}$. Once this is
done, we simply flip the state of $b$ again to drive the logical
bits back into the IFS. This procedure realizes a $CPHASE$ gate
between 2 encoded bits and puts them back in the IFS at the end.
Single bit operations (on the logical bits) can then be performed
to get other two bit gates like $CNOT$. A total number of 8 local
operations and 1 interaction are necessary, and the time required
is $\pi/16J_1$. This time is actually shorter than that needed in
an ordinary Ising model with switchable couplings, $\pi/4J_1$.
%since the encoded bits are ``more strongly'' coupled.

For completeness let us discuss briefly how the system can be
initialized in the IFS and how the states of the encoded bits can
be measured. If we apply a strong %(much stronger than the couplings)
global field in the $z$ direction to all the physical qubits, at
low temperatures all bits will line up with the field. Then
starting with the left most qubit, we can drive all the bits into
IFS by simply flipping the
state of one of $a$, $b$. %This can be done precisely, because the
%left most bit is coupled to the bit on its right only, who applies
%a field of $2J_1\hat{z}$ on it which can be balanced with a local
%field. Or we can give one of $a$, $b$ a strong $\pi$ pulse in the
%$x$ direction so that this small field can be disregarded. Once
%the left most bit is in IFS, it is decoupled from the rest of the
%qubits which we can continue to initialize.
To read out the state of the encoded bits, a measurement on its
$a$ or $b$ suffices.

With the procedures discussed above for diagonal interactions, we
can turn on and off the interactions between the logical qubits
without a physical switch. This is readily applicable in
superconducting quantum computation. Here, local fields can be
easily applied simply by changing the biases of the
superconducting qubits or applying ac fields. Typical times for
single bit gates range from hundreds of ps to tens of ns
\cite{ref:Charge1,ref:Flux2,ref:Charge-Switch1}, depending on the
type of the qubits and the choice of the parameters. Coherent
control of single superconducting qubits has been experimentally
realized \cite{ref:Coherent-Charge-Control,
ref:Coherent-Flux-Control}. %As discussed above, we may work in the
%weak interaction limit, choose the coupling strength such that
%single bit gates are fast compared to two bit operations.
With the fast operation speed \cite{ref:Coherent-Charge-Control}
and long decoherence time \cite{ref:Coherent-Flux-Control}
experimentally demonstrated, large scale superconducting quantum
computers can be constructed with the aid of IFS. %One practical
%concern is the uniform couplings required to satisfy the symmetry
%conditions for IFS (in Fig. \ref{fig:Diag}, couplings between
%physical qubits in neighboring logical bits should all be $J_1$).
%To compensate for possible fabrication errors or unknown couplings
%that could disturb the uniform couplings required (in Fig.
%\ref{fig:Diag}, couplings between physical qubits in neighboring
%logical bits should all be $J_1$), we can take advantage of
%developed variable coupling schemes
%\cite{ref:Flux2,ref:Flux-Switch} for the calibration of the
%system.
The uniform couplings required (in Fig. \ref{fig:Diag}, couplings
between physical qubits in neighboring logical bits should all be
$J_1$) are relatively easy to realize in superconducting designs,
as mutual inductances can be calculated and fabricated very
precisely. Were it necessary to compensate for fabrication
imperfections, simple schemes for minor adjustment of the coupling
are readily accomplished \cite{ref:Flux2}. This calibration step
can be very slow, so the leads used for calibration can be heavily
filtered to keep
out the noise. %Once the couplings have been calibrated we can then
%heavily filter the leads for the variable couplings to fix the
%couplings and keep out the noise.
For more detailed discussion on the application of our scheme in
superconducting quantum computing, see Ref. \cite{ref:MQC02}.

%One
%merit of our scheme is that it is fully scalable, so we have more
%freedom in choosing the parameters for the circuits, as opposed to
%the scheme discussed in \cite{ref:Charge-Switch1}, in which there
%is a tradeoff between the operation speed (determined by the
%typical eigenenergies of the qubit) and the number of qubits that
%can be incorporated in the system. For more detailed discussion on
%the application of our strategy in specific superconducting
%quantum computing schemes, see Ref. \cite{ref:MQC02}.

\begin{figure}[h]
%    \centering
%    \epsfig{file=aa1.eps, width=2.3in, height=1.2in}
    \includegraphics[width=3in, height=1.2in]{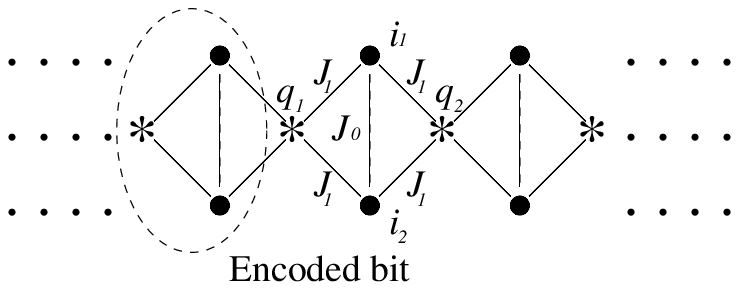}
    \caption{Architecture of the quantum computer for off diagonal interactions.
    Each logical bit consists of 3 physical bits, the information
    carrying qubit (the star) and an ``isolator'' (the dots
    connected with a dashed line). Couplings represented by the
    dashed lines are on in the initialization process. Those
    represented by the solid lines are on throughout the
    computation.}
    \label{fig:Off-Diag}
\end{figure}

We now turn to the case of off diagonal interactions. We will
focus on the isotropic and anisotropic exchange interactions,
$H_{int}=J_{xy}(\sigma_1^x \sigma_2^x +\sigma_1^y
\sigma_2^y)+J_z\sigma_1^z \sigma_2^z$, since these are most
frequently encountered in quantum computing proposals.
%Let us take the isotropic exchange interaction
%$H_{int}=J\vec{\sigma_1}\cdot \vec{\sigma_2}$ as the example, the
%anisotropic case can be treated in the same way (see below).
If both dissipation and decoherence are present, DFS requires 4
bits per encoded bit \cite{ref:DFS}. However as we discussed
earlier we have more control in constructing IFS, and it is
possible to use less resources. In our case 3 bits per logical bit
is enough. A few designs are possible, but let us consider the
architecture shown in Fig. \ref{fig:Off-Diag}. Here, each logical
bit contains 3 physical qubits. We have represented them with
stars and dots, not because they are physically distinct, but
because they play different roles. The stars are the information
carrying qubits, while the dots are ``isolators'' in the singlet
state $(|\uparrow_{i_1}\downarrow_{i_2}\rangle
-|\downarrow_{i_1}\uparrow_{i_2}\rangle)/\sqrt{2}$. To see the
origin of the name, we note that all the stars are coupled to its
neighboring dots with the same strength $J_1$ (meaning same
$J_{xy}$, $J_z$). Hence the interaction Hamiltonian between the
information carrier and its ``isolator'' is %$J_{xy} \vec{\sigma_q}
%\cdot (\vec{\sigma}_{i_1}+ \vec{\sigma}_{i_2})
$2J_{xy}(\sigma_q^+\Sigma_i^-+ \sigma_q^-\Sigma_i^+)
+J_z\sigma_q^z\Sigma_i^z$, where $\sigma^\pm=(\sigma_x\pm
i\sigma_y)/2$ and $\Sigma_i^{\pm,z}= \sigma_{i_1}^{\pm,z}
+\sigma_{i_2}^{\pm,z}$. Since the singlet state is annihilated by
the operators $\Sigma_i^{\pm,z}$, we see that if all the dots are
in the singlet state, the stars will be isolated from each other
and no phase exchange and state propagation will happen, hence the
name ``isolators''. The IFS is spanned by $|\uparrow_q\rangle
(|\uparrow_{i_1}\downarrow_{i_2}\rangle
-|\downarrow_{i_1}\uparrow_{i_2}\rangle)/\sqrt{2}$ and
$|\downarrow_q\rangle (|\uparrow_{i_1}\downarrow_{i_2}\rangle
-|\downarrow_{i_1}\uparrow_{i_2}\rangle)/\sqrt{2}$. To prepare the
isolators in the singlet state, we turn on all couplings between
the dots (the vertical dashed lines) while keep all couplings
between the stars and dots (the solid lines) off during the
initialization process. This is necessary to stop the propagation
of the states in the qubit array. Here we are assuming that
switching of the coupling is possible but hard and slow. Since
initialization and computation are subject to different
restrictions (initialization needs only be done once, and it does
not need to be done quickly), the global switches are used to
initialize the system. At low temperatures, the isolators will
then relax to the singlet which is the lowest energy eigenstate.
Once the initialization is done, we can then start the computation
by turning off all couplings represented by dashed lines and
turning on those represented by solid lines. According to our
assumption, these couplings will remain un-tuned throughout the
computation.

We see that single bit gates are trivial, we simply operate on the
stars directly as if they were not coupled to anything else. Fully
parallel operations are possible, thanks to the use of isolators.
Two bit gates are more complicated. For the convenience of
discussion, let us assume that we want to do a two bit operation
between the stars $q_1$ and $q_2$ in Fig. \ref{fig:Off-Diag} who
are separated by the isolator $i_1$, $i_2$. %For this purpose
%necessary single bit gates will be applied to $q_1$, $q_2$ and
%$i_1$, $i_2$. We want a gate sequence such that certain controlled
%operations between $q_1$, $q_2$ are realized while $i_1$, $i_2$
%are brought back to the singlet state at the end of the operation.
%A general approach for this problem is to write down a unitary
%gate on the full Hilbert space (16 dimensional) which performs the
%desired transformation on the IFS states and an arbitrary
%transformation on the rest of the Hilbert space, then try to
%synthesize this unitary gate with available resources, {\it i.e.},
%evolution under the interaction Hamiltonian
%$J_1(\vec{\sigma}_{q_1}+ \vec{\sigma}_{q_2}) \cdot
%(\vec{\sigma}_{i_1}+ \vec{\sigma}_{i_2}) $interspersed with local
%unitaries. Searching for such a unitary gate and its generating
%gate sequence is a nontrivial task because of the large dimension
%of the Hilbert space (because local gates will rotate the system
%states out of IFS), and numerical methods are necessary
%\cite{ref:No1bit}. Instead, we follow a simpler and more
%understandable approach of selective coupling, which allows us to
%select part of the entire interaction Hamiltonian. This is closely
%related to the problem of ``Hamiltonian simulation''
%\cite{ref:H-simu} and is based on the following observation.
In Fig. \ref{fig:Off-Diag}, we notice there is no coupling between
$q_1$ and $q_2$. In order to implement a controlled gate between
them, it is necessary to somehow transfer the state of the control
bit, say $q_1$, to the isolator which is coupled to the target bit
(or transfer the state of the target bit). One idea is to swap the
states of $q_1$ and $i_1$, perform a control gate between $i_1$
and $q_2$ and swap the states of $q_1$ and $i_1$ back. These steps
can be done, %if we can select specific parts of the interaction Hamiltonian.
if we can %synthesize unitary operators of the form
%$e^{-i(\sigma_{q_1}^x \sigma_{i_1}^x +\sigma_{q_1}^y
%\sigma_{i_1}^y)t}$
simulate dynamics generated by $\sigma_{q_1}^x \sigma_{i_1}^x
+\sigma_{q_1}^y \sigma_{i_1}^y$ (and similar for $i_1$ and $q_2$),
which can be used to generate swap and $CNOT$ gates
\cite{ref:Charge-Switch2}. This is possible with our available
resources, namely local unitaries and the interaction Hamiltonian
$H_{qi}=J_{xy}(\Sigma_q^x \Sigma_i^x +\Sigma_q^y
\Sigma_i^y)+J_z\Sigma_q^z \Sigma_i^z$ \cite{ref:Possibility}.
%since $\sigma_{q_1}^x
%\sigma_{i_1}^x$ and $\sigma_{q_1}^y \sigma_{i_1}^y$ are members of
%the algebra generated from $H_{qi}$ and local Pauli operators
%through commutation.
Here we discuss a method based on selective coupling. Note that
for a small time $t=\pi/(32NJ_{xy})$, where $N$ is some large
integer, we have $e^{-iH_{qi}t} \sigma_{q_1}^x \sigma_{q_2}^x
e^{-iH_{qi}t} \sigma_{q_1}^x \sigma_{q_2}^x= e^{-i 2J_{xy}t
\Sigma_{q}^x \Sigma_{i}^x} + o(t)$. This can be further used to
generate operators we want: $e^{-i 2J_{xy}t \Sigma_{q}^x
\Sigma_{i}^x} \sigma_{q_2}^z e^{-i 2J_{xy}t \Sigma_{q}^x
\Sigma_{i}^x} \sigma_{q_2}^z= e^{-i4J_{xy}t \sigma_{q_1}^x
\Sigma_i^x}$, $e^{-i4J_{xy}t \sigma_{q_1}^x \Sigma_i^x}
\sigma_{i_2}^z e^{-i4J_{xy}t \sigma_{q_1}^x \Sigma_i^x}
\sigma_{i_2}^z= e^{-i8J_{xy}t \sigma_{q_1}^x \sigma_{i_1}^x}$.
Similarly we can synthesize $e^{-i8J_{xy}t \sigma_{q_1}^y
\sigma_{i_1}^y}$. Repeating this procedure $N$ times, we then get
the transformation $e^{-i(\sigma_{q_1}^x \sigma_{i_1}^x
+\sigma_{q_1}^y \sigma_{i_1}^y)\pi/4}$, which is a swap gate
between $q_1$ and $i_1$ (and multiplication by $-i$ when their
states are different) \cite{ref:Charge-Switch2}. In order to
reduce the error, the number of repetition $N$ can be quite large
\cite{ref:H-Simu}, therefore many local gates ($44N$ in total) are
needed. The interaction time needed is $\pi/2J_{xy}$. The CNOT (or
CPHASE) gate between $i_1$ and $q_2$ after this swap operation,
and the operation to swap back the states of $i_1$ and $q_1$ can
be done by following the same procedure. The interaction times
required are $\pi/2J_{xy}$ each. Therefore the total interaction
time is $3\pi/2J_{xy}$, in comparison with $\pi/4J_{xy}$ which is
needed in a switchable XY model %(with zero switching time)
\cite{ref:XY-time}. This prescription verifies the possibility of
universal quantum computation in our current example. %though it is
%obviously not the most economical approach in terms of the number
%of local gates needed \cite{ref:Possibility}.
Finding physical systems to which our scheme discussed above can
apply, and a set of manipulations that allow to minimize the
complexity of the operation, is of further interest to us.

%In spite of the connection to
%Hamiltonian simulation, our problem is much easier than its most
%general case, because the scale of the ``simulation'' is
%restricted (4 bits involved), thanks to the isolators. It is also
%possible to get exact simulation with infrequent local unitaries,
%due to the particular form of the interaction Hamiltonian.

%We then discuss the procedure for a two bit operation following
%the above ideas. Selective coupling is a technique to ``reverse''
%the time evolution by taking advantage of the anti-commutativity
%between operators. For instance, as can be easily verified we have
%$\sigma^x e^{-ia\sigma^z} \sigma^x =e^{ia\sigma^z}$, $a$ being any
%operator commuting with $\sigma_x$. In our case, we first notice
%the following: $e^{-itH_{int}} \sigma_{q_1}^x \sigma_{q_2}^x
%e^{-itH_{int}} \sigma_{q_2}^x \sigma_{q_1}^x =e^{-i2tJ_1
%(\sigma_{q_1}^x +\sigma_{q_2}^x) (\sigma_{i_1}^x
%+\sigma_{i_2}^x)}$, where $H_{int}= J_1(\vec{\sigma}_{q_1}+
%\vec{\sigma}_{q_2}) \cdot (\vec{\sigma}_{i_1}+
%\vec{\sigma}_{i_2})$ is the interaction Hamiltonian of the system.
%This identity allows us to simulate exactly the Hamiltonian $J_1
%(\sigma_{q_1}^x +\sigma_{q_2}^x) (\sigma_{i_1}^x
%+\sigma_{i_2}^x)$, which can then be used to simulate exactly
%$\sigma_{q_1}^x \sigma_{i_1}^x$: $e^{-i2tJ_1 (\sigma_{q_1}^x
%+\sigma_{q_2}^x) (\sigma_{i_1}^x +\sigma_{i_2}^x)} \sigma$

In summary, we have devised a scheme for universal and scalable
quantum computation without the need to tune the couplings between
qubits. This relies on the idea of computing with logical bits
consisting of several physical qubits, which can be put in and
driven out of the IFS. We gave two examples of how universal
quantum computation can be done with our scheme. We emphasize that
our strategy can be adopted in forms and contexts other than the
examples discussed. %depending on the physical system.
With potential applications in superconducting quantum computation
and a few other cases, our scheme is likely to help simplify the
design and ease the operation of quantum computers.

Before we conclude, we should mention that the ``complimentary''
problem to ours, in which single bit operations are hard and
desired to be avoided, has been discussed \cite{ref:No1bit}. Our
scheme is much in the same spirit in the sense of using encoded
qubits for computation, but it is for a different purpose and it
has a closer relation to decoherence free subspaces (DFS). Other
schemes to reduce needed resources exist too. For instance, in
\cite{ref:Nolocal} a method of quantum computing without local
control of qubit-qubit interactions was studied. In
\cite{ref:Oneway}, the authors discussed how to do quantum
computing with only single bit measurements on a class of
entangled states, which are prepared by unitary evolution under
controllable Ising-type interactions.

%\begin{equation}
%\frac{d}{dt} \left (
%\begin{array}{c}
%s_x \\ s_y \\ s_z
%\end{array}
%\right ) =\left(
%\begin{array}{ccc}
%0 & -B_z & B_y \\
%B_z & 0 & -B_x \\
%-B_y & B_x & 0
%\end{array}
%\right ) \left(
%\begin{array}{c}
%s_x \\ s_y \\s_z
%\end{array} \right )
%\label{eq:matrix}
%\end{equation}

X. Zhou thanks L. Tian for helpful discussions. Work of X. Zhou
and M. J. Feldman was supported in part by AFOSR grant
F49620-01-1-0457 and funded under the DoD DURINT program and by
the ARDA. Z-W. Zhou and G-C. Guo were supported by National
Fundamental Research Program (2001CB309300), National Natural
Science Foundation of China, and the Innovation funds from Chinese
Academy of Sciences.

\end{document}